\documentclass[12pt]{article}
\usepackage{amsfonts,amssymb}

\newcommand{\nn}{\nonumber}
\newcommand{\ba}{\begin{eqnarray}}
\newcommand{\ea}{\end{eqnarray}}
\newcommand{\be}{\begin{equation}}
\newcommand{\ee}{\end{equation}}
\newcommand{\mbb }[1]{\mathbb #1}
\newcommand{\p}[1]{(\ref{#1})}

\begin{document}

\begin{center}
{\Large {\bf Hamiltonian structures of fermionic
}}\\[0.3cm]
{\Large {\bf two-dimensional  Toda lattice  hierarchies}}
\footnote{Contribution to the Proceedings of the
 International Workshop on Classical and Quantum Integrable Systems,
  Dubna, January 24--28, 2005.}\\[.7cm]

~\\

{\large  V.V. Gribanov\footnote{{\it E-mail address:}
gribanov@thsun1.jinr.ru}, V.G. Kadyshevsky,\footnote{{\it E-mail
address:} kadyshev@jinr.ru} and A.S. Sorin\footnote{{\it E-mail
address:} sorin@thsun1.jinr.ru}}

~\\

{{\em Bogoliubov Laboratory of Theoretical Physics,}}\\
{{\em Joint Institute for Nuclear Research,}}\\
{\em 141980 Dubna, Moscow Region, Russia}~\quad\\

\end{center}

~\\


\centerline{{\bf Abstract}} \noindent
  By exhibiting the corresponding Lax pair representations we propose a wide class
of integrable two-dimensional (2D) fermionic Toda lattice (TL)
hierarchies which includes the 2D $N=(2|2)$ and $N=(0|2)$
supersymmetric TL hierarchies as particular cases. We develop the
generalized graded R-matrix formalism using the generalized graded
bracket  on the space of graded operators with involution
generalizing the graded commutator in superalgebras, which allows
one to describe these hierarchies in the framework of the
Hamiltonian formalism and construct their first two Hamiltonian
structures. The first Hamiltonian structure is obtained  for both
bosonic and fermionic Lax operators while the second Hamiltonian
structure is established for bosonic Lax operators only.





\section{Introduction}

 The 2D TL hierarchy was first studied in
\cite{Mikh,UT}, and at present two different nontrivial
supersymmetric extensions of 2D TL are known. They are the
$N=(2|2)$ \cite{Olsh}-\cite{DLS} and $N=(0|2)$ \cite{DLS,KS1}
supersymmetric TL hierarchies that possess a different number of
supersymmetries and contain the $N=(2|2)$ and $N=(0|2)$ TL
equations as subsystems. Quite recently, the 2D generalized
fermionic TL equations have been introduced \cite{DDNS} and their
two reductions related to the $N=(2|2)$ and $N=(0|2)$
supersymmetric TL equations were considered. In the present paper,
we describe  a wide class of integrable two-dimensional fermionic
Toda lattice  hierarchies which includes the 2D $N=(2|2)$ and
$N=(0|2)$ supersymmetric TL hierarchies  as particular cases and
contains
 the 2D generalized
fermionic TL equations as a subsystem.

The Hamiltonian description of the 2D TL hierarchy has been
constructed only quite recently
 in the framework of the R-matrix approach  in \cite{Carlet}, where the
 new R-matrix associated with splitting of
 algebra given by a pair of difference operators was introduced.
 In the
present paper, we adapt this R-matrix to the case of $Z_2$-graded
operators and derive the bi-Hamiltonian structure of the 2D
fermionic $(K,M)$-TL hierarchy.

Remarkably,  in solving this problem the generalized graded
bracket \p{SK-bracket} on the space of graded  operators with an
involution finds its new application. This bracket was
 introduced in \cite{KS2}, where it was observed that the
$N=(1|1)$ supersymmetric 2D TL hierarchy had a  natural Lax-pair
representation in terms of this bracket which allowed one to
derive  the dispersionless $N=(1|1)$ 2D TL hierarchy and its  Lax
representation. In the present paper, the generalized graded
bracket is used to describe the 2D fermionic $(K,M)$-TL hierarchy
and define its two Hamiltonian structures.

The structure of this paper is as follows. In Sec.~2, we define
the space of the $Z_2$-graded difference operators with the
involution and recall  the generalized graded bracket  \cite{KS2}
and its properties. In Sec.~3, we give a theoretical background of
the $R$-matrix method generalized to the case  of the $Z_2$-graded
difference operators. We define the $R$-matrix on the associative
algebra {\sf g} of the $Z_2$-graded difference operators, derive
the graded modified  Yang-Baxter equation and using the
generalized graded bracket obtain two Poisson brackets for the
functionals on ${\sf g}^\dag={\sf g}$.
 Using these Poisson brackets one can define the
Hamiltonian equations that can equivalently be rewritten  in terms
of the Lax-pair representation. The basic results of Sec.~3 are
formulated as Theorem. In Sec.~4, using the generalized graded
bracket we propose a new 2D fermionic $(K,M)$-TL hierarchy in
terms of the Lax-pair representation and construct  the algebra of
its  flows. In Sec.~5, we apply the results of Sec.~3 to derive
the Hamiltonian structures of the  2D fermionic $(K,M)$-TL
hierarchy. In Sec.~6, we briefly summarize the main results
obtained in this paper and point out open problems.

\section{ Space of difference operators}
 In this section we define the space of difference operators
which will play an important role in our consideration. These
operators can be represented in the following general form:
 \ba  \label{O}
\mbb{O}_m=\sum\limits_{k=-\infty}^{\infty}f^{(m)}_{k,j}\ e^{(k-m)
\partial}, \ \ \  \ \ \ \ m, j \in \mathbb{Z},
 \ea
 parameterized
by the functions $f_{2k,j}^{(m)}$  ($f_{2k+1,j}^{(m)}$) which are
the $Z_2$-graded bosonic (fermionic) lattice fields with the
lattice index $j$ $(j\in\mbb{Z})$ and the Grassmann parity defined
by index $k$
\ba \label{f-grad} d_{f^{(m)}_{k,j}}=|k|\ \mbox{mod}\ 2. \nn
\ea
In what follows we suppose that the functions $f_{k,j}^{(m)}$ obey
the zero boundary conditions at infinity
\ba\label{bound} \lim_{j\rightarrow \pm \infty}f_{k,j}^{(m)}=0.
\ea
 $e^{k
\partial}$ is the shift operator whose action on the lattice fields
results into a discrete shift of a lattice index
 \ba \label{shift}
 e^{l \partial}f^{(m)}_{k,j}=f^{(m)}_{k,j+l}e^{l \partial}.
 \ea
The shift operator has  $Z_2$-parity defined as
\ba d_{e^{l\partial}}'=|l|\ \mbox{mod} \ 2.\nn \ea
The operators  $ \mbb{O}_m$ \p{O} admit the diagonal $Z_2$-parity
\ba\label{Z2-par}
d_{\mbb{O}_m}=d_{f_{k,j}^{(m)}}+d_{e^{(k-m)\partial}}'=|m|\
\mbox{mod} \ 2 \ea
 and the involution
$$
\mbb{O}^*_m=\sum\limits_{k=-\infty}^{\infty}(-1)^{k}f_{k,j}^{(m)}\
e^{(k-m)
\partial}.\nn $$

 In what follows we also need the projections
 of the operators $\mbb{O}_m$ defined as
\ba  ( \mbb{O}_m)_{\leqslant p}=\sum\limits_{k\leqslant
p+m}f_{k,j}\ e^{(k-m)
\partial},\ \ \ \
 (\mbb{O}_m)_{\geqslant p}=\sum\limits_{k\geqslant p+m}f_{k,j}\
 e^{(k-m)
\partial}\nn
 \ea
and we will use the usual notation for the projections
$(\mbb{O}_m)_+:=(\mbb{O}_m)_{\geqslant 0}$ and
$(\mbb{O}_m)_-:=(\mbb{O}_m)_{< 0}$.
 Note that $e^{l\partial}$  is a conventional form for the
shift operators defined in terms of infinite-dimensional matrices
 $(e^{l\partial})_{i,j}\equiv\delta_{i,j-l}$, and
there is an isomorphism between operators \p{O} and
infinite-dimensional matrices (see e.g. \cite{AB})
\ba \mbb{O}_m=\sum\limits_{k=-\infty}^{\infty}f_{k,j}^{(m)}\
e^{(k-m)
\partial}\ \ \to \ \ {(\mbb{O}_m)}_{j,i}\equiv\sum\limits_{k=-\infty}^{\infty} f_{k,j}^{(m)}\
\delta_{j,i-k+m}.\nn \ea

In the operator space \p{O} one can extract two subspaces which
are of great importance  in our further consideration
\ba  \label{O+}
\mbb{O}^+_{K_1}&=&\sum\limits_{k=0}^{\infty}f_{k,j}\ e^{(K_1-k)
\partial},\ \ \ \  \ K_1\in\mathbb{N},\\
 \label{O-} \mbb{O}^-_{K_2}&=&\sum\limits_{k=0}^{\infty}f_{k,j}\
e^{(k-K_2)
\partial},\ \ \ \  \ K_2\in\mathbb{N}.
%
 \ea
 \looseness=-1
The operators of  the subspaces $ \mbb{O}^+_{K_1}$ and
$\mbb{O}^-_{K_2}$  form associative algebras with the
multiplication \p{shift}. Using this fact we define on these
subspaces the generalized graded  algebra with the bracket
\cite{KS2}
\ba\label{SK-bracket}  [{\mathbb O}, \widetilde{ \mathbb O}
\}:={\mathbb O}\ \widetilde{\mathbb O}
 - (-1)^
 {d_{{\mathbb O}
 \vphantom{\widetilde{\mathbb O}} }
 d_{\widetilde{\mathbb O}}}~{\widetilde{\mathbb O}}^{*(d_{{\mathbb
O}\vphantom{\widetilde{\mathbb O}}})}~ {{\mathbb
O}}^{*(d_{\widetilde{\mathbb O}})}, \ea
 where the operators ${{\mathbb O}}$  and  ${\widetilde{\mathbb O}}$ belong to the
 subspaces $\mbb{O}^+_{K_1}\ (\mbb{O}^-_{K_2})$, and
${{\mathbb O}}^{*(m)}$  denotes the $m$-fold action of the
involution $*$ on the operator ${\mathbb O}$,  (${{\mathbb
O}}^{*(2)}={\mathbb O}$). Bracket \p{SK-bracket} generalizes  the
(anti)commutator in superalgebras and satisfies the following
properties \cite{KS2}:

symmetry
 \ba \label{symSK} [ {\mathbb O}, \widetilde{ \mathbb O}
\}=
 - (-1)^
 {d_{{\mathbb O}
 \vphantom{\widetilde{\mathbb O}} }
 d_{\widetilde{\mathbb O}}}~[{\widetilde{\mathbb O}}^{*(d_{{\mathbb
O}\vphantom{\widetilde{\mathbb O}}})},  {{\mathbb
O}}^{*(d_{\widetilde{\mathbb O}})}\}, \ea

 derivation
 \ba \label{derSK}
 [{\mathbb O}, \widetilde{ \mathbb O}\, \widehat{ \mathbb O}
\}=[ {\mathbb O}, \widetilde{ \mathbb O}\}\, \widehat{ \mathbb O}
 + (-1)^{d_{{\mathbb O}
 \vphantom{\widetilde{\mathbb O}} } d_{\widetilde{\mathbb O}}}\,
 {\widetilde{\mathbb
O}}^{*(d_{{\mathbb O}\vphantom{\widetilde{\mathbb O}}})} [{\mathbb
O}^{*(d_{\widetilde{\mathbb O}})}, {\widehat{\mathbb O}}\},\ea

 and  Jacobi identity
\ba\label{SK-Jacobi} (-1)^{d_{{\mathbb O}
 \vphantom{\widetilde{\mathbb O}} } d_{\widehat{\mathbb O}\vphantom{\widetilde{\mathbb
 O}}}}\,
 [[{{\mathbb O}},\, {\widetilde{\mathbb O}}^{*(d_{{\mathbb
O}\vphantom{\widetilde{\mathbb O}}})}\},\,  {\widehat{\mathbb
O}}^{*(d_{{\mathbb O}\vphantom{\widetilde{\mathbb O}}}+
d_{\widetilde{\mathbb O}}) \vphantom{\widetilde{\mathbb O}}}\}
 +(-1)^{d_{\widetilde{{\mathbb O}}
 \vphantom{\widetilde{\mathbb O}} } d_{{\mathbb O}\vphantom{\widetilde{\mathbb
 O}}}}\,
 [[{\widetilde{\mathbb O}},\, {\widehat{\mathbb O}}^{*(d_{\widetilde{{\mathbb
O}}\vphantom{\widetilde{\mathbb O}}})}\},\,
 {{\mathbb O}}^{*(d_{\widetilde{{\mathbb O}}\vphantom{\widetilde{\mathbb
O}}}+d_{\widehat{\mathbb O}}) \vphantom{\widetilde{\mathbb
O}}}\}&&\nn\\
 +\ \  (-1)^{d_{\widehat{\mathbb O}
 \vphantom{\widetilde{\mathbb O}} } d_{\widetilde{\mathbb O}\vphantom{\widetilde{\mathbb
 O}}}}\,
[[{\widehat{\mathbb O}},\, {{\mathbb O}}^{*(d_{\widehat{\mathbb
O}\vphantom{\widetilde{\mathbb O}}})}\},\,
 {\widetilde{\mathbb O}}^{*(d_{\widehat{{\mathbb O}}\vphantom{\widetilde{\mathbb
O}}}+d_{{\mathbb O}}) \vphantom{\widetilde{\mathbb O}}}\}
 &=&0.
\ea

For the operators $\mathbb{O}_m$ \p{O} we define the supertrace
\ba\label{supertr}
 str \mathbb O=\sum\limits_{j=-\infty}^{\infty}(-1)^j
f^{(m)}_{m,j}. \ea
One can easily verify that the main property of supertraces
$str [ {\mathbb O}, \widetilde{\mathbb O} \}=0$
 is indeed satisfied for the case of the
generalized graded bracket  \p{SK-bracket} if the functions
entering into operators \p{O} obey the zero boundary condition at
infinity \p{bound}.

\section{R-matrix formalism}
In this section, we develop  a theoretical background of the
R-matrix method adapted to the case of the operator space \p{O}.

 Let {\sf g} be an associative algebra of the operators from the space \p{O} with
the invariant non-degenerate inner product
\ba \label{InProd}
<\mbb{O},\widetilde{\mbb{O}}>=str(\mbb{O}\,\widetilde{\mbb{O}})\nn
\ea
using which one can identify the algebra {\sf g} with its dual
{\sf g${}^\dag$}.  We set the following Poisson bracket:
\ba \label{pb1} \{f,g\}(\mbb{O})=-<\mbb{O},[\nabla g,(\nabla
f)^{*(\nabla d_{g})}\} >, \ea
where $f, g$ are functionals on {\sf g},
and $\nabla f$ and  $\nabla g$ are their gradients at the point
$\mbb{O}$
 which are
related with $f,g $ through  the inner product
\ba \frac{\partial f(\mbb{O}+\epsilon \delta
\mbb{O})}{\partial\epsilon}{\Biggr |}_{\epsilon=0} =\ <\delta
\mbb{O}, \nabla f(\mbb{O})>. \nn\ea
%
%
%
Note that the proper properties of the Poisson bracket \p{pb1}
follow from the properties (\ref{symSK}--\ref{SK-Jacobi}) of the
generalized bracket \p{SK-bracket} and are strictly determined by
the $Z_2$-parity of the operator $\mathbb{O}$. Thus, one has
 symmetry
 \ba\label{sym}
\{f,g\}&=&-(-1)^{(d_f+d_\mathbb{O})(d_g+d_\mathbb{O})}\{g,f\}, \ea
 derivation
\ba \label{der} \{f,g h\}&=&
\{f,g\}h+(-1)^{d_g (d_f+d_\mathbb{O})} g\{f,h\},
 \ea
%
%
and Jacobi identity
 \ba\label{Jac}
(-1)^{(d_f+d_\mathbb{O})(d_h+d_\mathbb{O})}\{\{f,g\},h\}
+(-1)^{(d_g+d_\mathbb{O})(d_f+d_\mathbb{O})}\{\{g,h\},f\}\nn\\
 +(-1)^{(d_h+d_\mathbb{O})(d_g+d_\mathbb{O})}\{\{h,f\},g\}&=&0. \ea
Therefore, for the even operator  $\mathbb{O}$ one has usual
(even) $Z_2$-graded Poisson bracket, while for the operators with
odd diagonal parity $d_\mathbb{O}$ \p{Z2-par}  eq. \p{pb1} defines
odd $Z_2$-graded Poisson bracket (antibracket).

Having defined the   Poisson bracket   we proceed with the search
for the hierarchy of  flows generated by this  bracket using
Hamiltonians. Therefore, we need to determine an infinite set of
functionals which should be in involution to play the role of
Hamiltonians. For Poisson bracket \p{pb1} one can find
 an infinite set of Hamiltonians in a rather standard way
\ba \label{hamB} H_k=\frac1k
str\mbb{O}^k_*=\frac1k\sum_{i=-\infty}^\infty(-1)^i
f_{km,i}^{(km)}, \ea
where $ \mbb{O}^k_*$ is defined as
\ba \label{compLax}(\mbb{O})^{2k}_*:=
(\mbb{O}^{*(d_\mbb{O})}\mbb{O})^k, \ \ \ (\mbb{O})^{2k+1}_*:=
\mbb{O}\ (\mbb{O})^{2k}_* .\ea
 For the odd operators $\mbb{O}$ eq. \p{hamB} defines only
fermionic nonzero functionals $H_{2k+1}$, since in this case  even
powers of the operators $\mbb{O}$ have the following
representation:
\ba \label{evenH} d_{\mathbb{O}}=1:\ \ \ \ (\mathbb{O})^{2k}_*=
(1/2[(\mathbb{O})^{*},\mathbb{O}\})^k \equiv
1/2[({(\mathbb{O})^{2k-1}_*})^{*},\mathbb{O}\}  \ea
and all the bosonic Hamiltonians are trivial ($H_{2k}=0$ ) like
the supertrace of the generalized graded bracket.

  The  functionals \p{hamB}
 are obviously in involution but produce a trivial dynamics.
 Actually, the functionals $H_k$
\p{hamB}  are the Casimir operators of the Poisson bracket
\p{pb1}, so the Poisson bracket of $H_k$ with any other functional
is equal to zero as an output (due to the relation $\nabla
H_{k+1}=\mbb{O}_*^k$). Nevertheless, it is possible to
 modify  the Poisson bracket \p{pb1} in such a way that the new Poisson
bracket would produce nontrivial  equations of motion using the
same Hamiltonians \p{hamB}  and these Hamiltonians  are in
involution with respect to the modified Poisson bracket as well.
Let us introduce the modified generalized  graded bracket on the
space \p{O}
\ba\label{SKmod}
[\mbb{O},\widetilde{\mbb{O}}\}_R:=[R(\mbb{O}),\widetilde{\mbb{O}}\}+[\mbb{O},R(\widetilde{\mbb{O}})\},
\ea
where the $R$-matrix
is a linear map $R$: {\sf g $\to$ g}  such that the  bracket
\p{SKmod} satisfies  the properties
(\ref{symSK}--\ref{SK-Jacobi}). One can verify that the Jacobi
identities \p{SK-Jacobi} for the bracket \p{SKmod} can
equivalently be
 rewritten in terms of the generalized graded bracket
\p{SK-bracket}
\ba  (-1)^{d_{{\mathbb O}
 \vphantom{\widetilde{\mathbb O}} } d_{\widehat{\mathbb O}\vphantom{\widetilde{\mathbb
 O}}}}\,
 [[{{\mathbb O}},\, {\widetilde{\mathbb O}}^{*(d_{{\mathbb
O}\vphantom{\widetilde{\mathbb O}}})}\}_R,\, {\widehat{\mathbb
O}}^{*(d_{{\mathbb O}\vphantom{\widetilde{\mathbb O}}}+
d_{\widetilde{\mathbb O}}) \vphantom{\widetilde{\mathbb
O}}}\}_R+\mbox{cycle
perm.}=~~~~~~~~~~~~~~~~~~~~~~~~~~~~~~\nn\\
(-1)^{d_{{\mathbb O}
 \vphantom{\widetilde{\mathbb O}} } d_{\widehat{\mathbb O}\vphantom{\widetilde{\mathbb
 O}}}}\, [R( [{{\mathbb O}},\, {\widetilde{\mathbb O}}^{*(d_{{\mathbb
O}\vphantom{\widetilde{\mathbb O}}})}\}_R )- [R({{\mathbb O}}),\,
R({\widetilde{\mathbb O}}^{*(d_{{\mathbb
O}\vphantom{\widetilde{\mathbb O}}})})\},{\widehat{\mathbb
O}}^{*(d_{{\mathbb O}\vphantom{\widetilde{\mathbb O}}}+
d_{\widetilde{\mathbb O}}) \vphantom{\widetilde{\mathbb
O}}}\}+\mbox{cycle perm.}=0.\nn
 \ea
Thus, one can conclude that a sufficient condition for $R$ to be
the $R$-matrix is the validity of the following equation:
\ba \label{YB} R([\mbb{O},\widetilde{\mbb{O}}\}_R)-
[R(\mbb{O}),R(\widetilde{\mbb{O}})\}=\alpha
[\mbb{O},\widetilde{\mbb{O}}\}, \ea
where $\alpha$ is an arbitrary constant. Following the terminology
of \cite{STSh} we call
 eq. \p{YB} the graded  modified  Yang-Baxter equation.
 Eq. \p{YB} is the
generalization of the    graded modified classical Yang-Baxter
equation discussed in paper \cite{Yung} for the space of graded
operators \p{O}.

With the new bracket \p{SKmod}  one can define the corresponding
new Poisson bracket on dual~${\sf g}^\dag$
\ba \label{PB1} \{f,g\}_1(\mbb{O})&=&-1/2<\mbb{O},[\nabla
g,(\nabla f)^{*( d_{\nabla g})}\}_R >.
%
 \ea
%
%
%
With respect to the dependence of the r.h.s of \p{PB1} on the
point $\mathbb{O}$ this is a linear bracket. Without going into
details we introduce also bi-linear bracket for bosonic graded
operators $\mathbb{O}_B$  ($d_{\mathbb{O}_B}=0$) as follows:
\ba\label{PB2} \{f,g\}_2(\mathbb{O}_B)&=&-1/4<[\mathbb{O}_B,\nabla
g\}R((\nabla f)^{*(d_{\nabla g})} \mathbb{O}_B^{*(d_{\nabla f}+d_{\nabla g})}\nn\\
&+&\mathbb{O}_B^{*(d_{\nabla  g})}(\nabla f)^{*(d_{\nabla g})})
-R(\nabla g \mathbb{O}_B^{*(d_{\nabla g})}+\mathbb{O}_B\nabla
g)[\mathbb{O}_B^{*(d_{\nabla  g})},(\nabla f)^{*( d_{\nabla
g})}\}>.~~~~  \ea
We did not succeed in constructing the bi-linear bracket for the
case of fermionic operators $\mathbb{O}_F$ ($d_{\mathbb{O}_F}=1$).
The bracket \p{PB1} is obviously the  Poisson bracket if $R$ is an
$R$-matrix on ${\sf g}$. The bi-linear bracket \p{PB2} becomes
Poisson bracket under more rigorous  constraints which can be
found in the following
%
%

{\bf ~~Theorem.}\ a) Linear bracket \p{PB1} is the Poisson
bracket if $R$ obeys the graded modified Yang-Baxter equation \p{YB};\\
b) the bi-linear bracket \p{PB2} is the Poisson bracket if $R$ and
its skew-symmetric part $1/2(R-R^\dag)$ obey the graded  modified
Yang-Baxter equation \p{YB} with the same $\alpha$, where the
adjoint operator $R^\dag$ acts on the dual $\sf g^\dag$
$$<\mathbb{O},R(\widetilde{\mathbb{O}})>=<R^\dag(\mathbb{O}),\widetilde{\mathbb{O}}>;$$
%
%
c) if $\mathbb{O}=\mathbb{O}_B$, then these two Poisson brackets
are compatible
\ba \{f,g\}_2(\mathbb{O}_B+b)=\{f,g\}_2(\mathbb{O}_B)+b
\{f,g\}_1(\mathbb{O}_B);\nn \ea
%
%
%
d) the  Casimir operators $H_{k}$ \p{hamB}  of the bracket \p{pb1}
are in involution with respect to both linear \p{PB1} and
bi-linear \p{PB2} Poisson brackets;\\
e) the Hamiltonians $H_{k}\neq 0$  \p{hamB} generate evolution
equations
\ba\label{Lax-R}
\partial_k \mathbb{O}&=&\{H_{k+1},\mathbb{O}\}_1=1/2 [R((\nabla H_{k+1})^{*(d_\mathbb{O})}),\mathbb{O}\},\nn\\
\partial_k \mathbb{O}_B&=&\{H_k,\mathbb{O}_B\}_2=1/4 [R(\nabla H_k \mathbb{O}_B+\mathbb{O}_B\nabla
H_k),\mathbb{O}_B\}\nn\ea
via the brackets \p{PB1} and \p{PB2}, respectively, which connect
the Lax-pair and Hamiltonian representations. ~~
$\blacktriangledown$

Note that in the case when  the shift operators and functions
parameterizing the difference operators $\mathbb{O}$ \p{O} have
even $Z_2$-parity the similar Theorem was discussed in
\cite{STSh,OR,LCP,Carlet}.

\section{ 2D  fermionic $(K,M)$-Toda lattice hierarchy}
In this section, we introduce the two-dimensional fermionic
$(K,M)$-Toda lattice hierarchy in terms of the Lax-pair
representation.

 Let us consider two difference operators $L^+_K$ and $L^-_M$
 \ba\label{LaxOp} L^+_K=\sum\limits_{k=0}^\infty u_{k,i}
e^{(K-k)\partial}, \ \ \ \ \ \ \ L^-_M=\sum\limits_{k=0}^\infty
v_{k,i} e^{(k-M)\partial},
 \ea
 which obviously belong to the spaces \p{O+} and \p{O-},
respectively. The lattice fields   and the shift operator entering
into these operators have the following length dimensions:
$[u_{k,i}]=-\frac{1}{2} k$,
 $[v_{k,i}]=\frac{1}{2}(k-K-M)$  and $[e^{k\partial}]=-\frac{1}{2} k$,
 respectively, so  operators
 \p{LaxOp} are of equal length dimension,
 $[L^+_K]=[L^-_M]=-\frac{1}{2} K$.
%
%
%
%
The dynamics of the fields $u_{k,i},v_{k,i}$ are governed by the
Lax equations expressed in terms of the generalized bracket
\p{SK-bracket} \cite{KS2}
\begin{eqnarray}\label{LaxEq}
\ D_s^\pm L^\alpha_{\Omega^\alpha}&=&\mp \alpha (-1)^{s
\Omega^\alpha \Omega^\pm
}[(((L^\pm_{\Omega^\pm})_*^s)_{-\alpha})^{*(\Omega^\alpha)},
L^\alpha_{\Omega^\alpha}\},\nn\\
 &&~ \alpha=+,-,\ \ \ \Omega^+=K,\
\ \ \Omega^-=M, \ \ \ s \in \mathbb{N},
 \ea
%
%
where $D^\pm_s$  are evolution derivatives with the $Z_2$-parity
defined as
$$ d_{D^+_s}=sK\ \mbox{ mod } \ 2, \ \ \ \ \ d_{D^-_s}=sM\ \mbox{
mod } \ 2 $$ and the length dimension $[D^+_s]=[D^-_s]= - sK/2.$
The Lax equations \p{LaxEq} generate  non-Abelian (super)algebra
of flows of the 2D fermionic $(K,M)$-TL hierarchy
\ba [D^+_s,D^+_p\}=(1-(-1)^{spK})D^+_{s+p},\ \ \ \
\bigl[D^-_s,D^-_p\}   =  (1-(-1)^{spM})D^-_{s+p},\ \ \ \
 \bigl[ D^+_s,D^-_p\} =   0. \nn
 \ea
The composite  operators $(L^+_K)_*^s$ and  $(L^-_M)_*^s$ entering
into the
 Lax equations \p{LaxEq} are defined by eq. \p{compLax}
  and  also belong to the spaces
\p{O+} and \p{O-}, respectively,
\ba\label{LaxPower} (L^+_K)^r_*:=\sum\limits_{k=0}^{\infty}
u_{k,i}^{(r)} e^{(rK-k)\partial}, \ \ \ \ \ \ \
(L^-_M)^r_*:=\sum\limits_{k=0}^\infty v_{k,i}^{(r)}
e^{(k-rM)\partial}.~~~~~~~\nn \ea
Here $u_{k,i}^{(r)}$ and $v_{k,i}^{(r)}$ are functionals of the
original fields
 and there are the  following recursion relations for them
\ba\label{CompFields} u_{p,i}^{(r+1)}&=&\sum\limits_{k=0}^p
(-1)^{k K} u_{k,i}^{(r)}u_{p-k,i-k+rK},\ \ \ \ u_{p,i}^{(1)}=u_{p,i},\nn\\
v_{p,i}^{(r+1)}&=&\sum\limits_{k=0}^{p}(-1)^{k M}
v_{k,i}^{(r)}v_{p-k,i+k-rM}, \ \ \ \ v_{p,i}^{(1)}=v_{p,i}. \nn\ea
Now using the Lax representation \p{LaxEq} and relations \p{derSK}
and \p{compLax} one can derive the equations of motion for the
composite Lax operators
\begin{eqnarray}\label{LaxEqCom}
\ D_s^\pm (L^\alpha_{\Omega^\alpha})^r_*&=&\mp \alpha (-1)^{s  r
\Omega^\alpha \Omega^\pm
}[(((L^\pm_{\Omega^\pm})_*^s)_{-\alpha})^{*(r\Omega^\alpha)},
(L^\alpha_{\Omega^\alpha})^r_*\}.
 \ea

%
%
%
 The Lax-pair representation  (\ref{LaxEqCom}) generates the
following equations for the functionals
$u_{k,i}^{(r)},v_{k,i}^{(r)}$:
\ba \label{Eqs1} D^+_s u_{k,i}^{(r)}&=&\sum\limits_{p=1}^{k}
((-1)^{rpK+1}u_{p+sK,i}^{(s)}u_{k-p,i-p}^{(r)}
+(-1)^{(k+p)sK}u_{k-p,i}^{(r)}u_{p+sK,i+p-k+rK}^{(s)}),\nn\\
D^-_s u_{k,i}^{(r)}&=&\sum\limits_{p=0}^{sM-1}
((-1)^{(sM+p)rK}v_{p,i}^{(s)}u_{p+k-sM,i+p-sM}^{(r)}
-(-1)^{(k+p+1)sM}u_{p+k-sM,i}^{(r)}v_{p,i-p-k+sM+rK}^{(s)}),\nn\\
D^+_s v_{k,i}^{(r)}&=&\sum\limits_{p=0}^{sK}
((-1)^{(sK+p)rM}u_{p,i}^{(s)}v_{p+k-sK,i-p+sK}^{(r)}
-(-1)^{(k+p+1)sK}v_{p+k-sK,i}^{(r)}u_{p,i+p+k-sK-rM}^{(s)}),\nn\\
D^-_s v_{k,i}^{(r)}&=&\sum\limits_{p=0}^{k}
((-1)^{rpM+1}v_{p+sM,i}^{(s)}v_{k-p,i+p}^{(r)}
+(-1)^{(k+p)sM}v_{k-p,i}^{(r)}v_{p+sM,i+k-p-rM}^{(s)}).
\ea
It is assumed that in the right-hand side of  eqs. (\ref{Eqs1})
 all the
functionals $u_{k,i}^{(r)}$ $v_{k,i}^{(r)}$ with $k<0$  should be
set equal to zero.

One can demonstrate that all known up to now fermionic 2D Toda
lattice equations \cite{Ik}-\cite{DLS}  can be reproduced from the
system of equations (\ref{Eqs1}) as subsystems with additional
reduction constraints imposed. We call equations \p{LaxEq} the 2D
 fermionic (K,M)-Toda lattice hierarchy.

\section{Bi-Hamiltonian structure of 2D fermionic $(K,M)$-TL hierarchy}
In this section, we apply the R-matrix approach to build the
bi-Hamiltonian structure of the 2D  fermionic (K,M)-TL hierarchy.
This hierarchy is associated with two Lax operators \p{LaxOp}
belonging to the operator space (\ref{O+}-\ref{O-}). Following
\cite{Carlet} we consider the associative algebra on the space of
the direct sum of two
 difference operators
\ba \label{O2} {\sf g}:= \mathbb{O}^+_{K_1}\oplus
\mathbb{O}^-_{K_2}. \ea
However, in contrast  to the case of pure bosonic 2D TL hierarchy,
the difference operators in the direct sum \p{O2} can be of both
opposite and  equal  diagonal $Z_2$-parity. It turns out that the
Poisson bracket can correctly be  defined only for the latter
case.  In what follows we restrict ourselves to the case when both
operators in {\sf g} \p{O2} have the same diagonal parity.

 We denote  $(x^+,x^-)$  elements of such algebra ${\sf g}={\sf
g}^\dag$ with the product
\ba\label{product}
(x^+_1,x^-_1)\cdot(x^+_2,x^-_2)=(x^+_1x^+_2,x^-_1x^-_2), \ea
and define the inner product  as follows:
 \ba \label{Ipr2}
<(x^+, x^-)>:=str(x^++x^-), \ea
 where $x^+\in \mathbb{O}^+_{K_1}$, $x^-\in\mathbb{O}^-_{K_2}$.
%
%
 Using this definition we set the Poisson brackets
 to be
\ba \label{pb2D} \{f_1,f_2\}= <(\mathbb{O}^+_{K_1},
\mathbb{O}^-_{K_2}),[\nabla f_1,\nabla f_2\}^\oplus>, \ea
where $$[\nabla f_1,\nabla f_2\}^\oplus:= ([\nabla f_1^+,(\nabla
f_2^+)^{*(d_{\nabla f_1^+})}\},[\nabla f_1^-,(\nabla
f_2^-)^{*(d_{\nabla f_1^-})}\}),$$
$f_k$ are functionals on {\sf g} \p{O2}, and $\nabla
f_k[(\mathbb{O}^+_{K_1},\mathbb{O}^-_{K_2})]=(\nabla f^+_k,\nabla
f^-_k)$ are their gradients which can be  found from the
definition
 \ba \frac{\partial f_k[(\mathbb{O}^+_{K_1},\mathbb{O}^-_{K_2})+\epsilon (\delta \mathbb{O}^+_{K_1},
\delta \mathbb{O}^-_{K_2})] }{\partial\epsilon}{\Biggr
|}_{\epsilon=0} &=& <(\delta \mathbb{O}^+_{K_1}, \delta
\mathbb{O}^-_{K_2}), (\nabla f^+_k,\nabla
f^-_k)>\nn\\
&=&<\delta \mathbb{O}^+_{K_1},\nabla f^+_k>+<\delta
\mathbb{O}^-_{K_2},\nabla f^-_k>.~~~~~~~~~ \nn\ea
 In order to
obtain nontrivial Hamiltonian dynamics,  one needs to modify the
brackets \p{pb2D} applying the $R$-matrix
\ba [\nabla f_1,\nabla f_2\}^\oplus\ \ \longrightarrow \ \ [\nabla
f_1,\nabla f_2\}^\oplus_R=[R(\nabla f_1),\nabla f_2\}^\oplus+
[\nabla f_1,R(\nabla f_2)\}^\oplus. \nn\ea
 The $R$-matrix acts on the space \p{O2} in the
nontrivial way  and mixes up the elements from two subalgebras in
the direct sum with each other
\ba \label{2dR} R(x^+,x^-)=(x^+_+-x^+_-+2x^-_-,x_-^-
-x^-_++2x^+_+) \ea
which is a crucial point of the $R$-matrix approach in the
two-dimensional case \cite{Carlet}. This $R$-matrix satisfies  the
graded modified  Yang-Baxter equation
\ba \label{YB2D}R([(x^+,x^-),(y^+,y^-)\}_R)-
[R(x^+,x^-),R(y^+,y^-)\}=\alpha [(x^+,x^-),(y^+,y^-)\} \ea
with $\alpha =1$ and allows one to find two compatible Poisson
structures and rewrite Lax-pair representation \p{LaxEq} in the
Hamiltonian form. The direct verification by substitution in
\p{YB2D} shows that the skew-symmetric part
\ba 1/2 (R(x^+,x^-)-R^\dag
(x^+,x^-))=(x^+_{>0}-x^+_{<0}-x^-_0,x_{<0}^- -x^-_{>0}+x^+_0)
\nn\ea
also satisfies the   graded modified  Yang-Baxter equation
\p{YB2D}. Therefore, by Theorem of section~3 there exist two
Poisson structures on {\sf g} \p{O2}.

Using  eqs. (\ref{PB1}--\ref{PB2}), \p{product}, \p{2dR} and cyclic
permutations inside the supertrace \p{supertr} we obtain the
following general form of the first and second Poisson brackets:
\ba\label{PB-2d} \{f,g\}_i= <P^+_i(\nabla g^+,\nabla g^-),(\nabla
f^+)^{*(d_{\nabla g})}>+<P^-_i(\nabla g^+,\nabla g^-),(\nabla
f^-)^{*(d_{\nabla g})}>, \ea
where $i=1,2$ and $d_{\nabla g}:=d_{\nabla g^+}=d_{\nabla g^-}$.
 The Poisson tensors in eq. \p{PB-2d}  are found for any values of $(K,M)$
 for the first Hamiltonian structure
\ba P_1^+(\nabla g^+,\nabla g^-)\ =\ [(\nabla g^-_-- \nabla
g^+_-)^{*(K)},(L^+_K)^{*(d_{\nabla g})}\}
 -([L^+_K,\nabla g^+\}+[L^-_M,\nabla g^-\})_{\leqslant 0},\nn\\
P_1^-(\nabla g^+,\nabla g^-)\ =\ [(\nabla g^+_+- \nabla
g^-_+)^{*(M)},(L^-_M)^{*(d_{\nabla g})}\}
-([L^+_K,\nabla g^+\}+[L^-_M,\nabla g^-\})_{> 0},\nn\ea
while for the second Hamiltonian structure we constructed the
explicit   expression of the Poisson tensors  for even values of
$(K,M)$ only
\ba P_2^+(\nabla g^+,\nabla g^-)\!\!\! &=&\!\!\! 1/2\Bigr
([(\nabla g^-(L^-_M)^{*(d_g)}+L^-_M\nabla g^-
-\nabla g^+(L^+_K)^{*(d_g)}
-L^+_K\nabla g^+)_-,(L^+_K)^{*(d_g)}\}\nn\\
\!\!\! & -&\!\!\! L^+_K\ ([L^+_K,\nabla g^+\}+[L^-_M,\nabla
g^-\})_{\leqslant 0}
-([L^+_K,\nabla g^+\}+[L^-_M,\nabla g^-\})_{\leqslant 0}\ (L^+_K)^{*(d_g)}\Bigl),\nn\\
P_2^-(\nabla g^+,\nabla g^-)\!\!\!&=&\!\!\!1/2\Bigr ([(\nabla
g^+(L^+_K)^{*(d_g)}+L^+_K\nabla g^+
-\nabla g^-(L^-_M)^{*(d_g)}
-L^-_M\nabla g^-)_+,(L^-_M)^{*(d_g)}\}\nn\\
\!\!\!&-&\!\!\!L^-_M\ ([L^+_K,\nabla g^+\}+[L^-_M,\nabla
g^-\})_{>0}
-([L^+_K,\nabla g^+\} +[L^-_M,\nabla g^-\})_{> 0}\
(L^-_M)^{*(d_g)}\Bigl).\nn
 \ea
The Poisson brackets for the functions $u_{n,i}$ and $v_{n,i}$
parameterizing the Lax operators \p{LaxOp} can  explicitly  be
derived from \p{PB-2d} if one takes into account  that
\ba\label{grad-uv} \nabla u_{n,\xi}&\equiv&(\nabla
u_{n,\xi}^+,\nabla u_{n,\xi}^-)=(e^{(n-K)\partial}(-1)^i
\delta_{i,\xi},0),\nn\\
 \nabla v_{n,\xi}&\equiv&(\nabla v_{n,\xi}^+,\nabla
v_{n,\xi}^-)=(0,e^{(M-n)\partial}(-1)^i \delta_{i,\xi}).\nn\ea
In such  a way one can obtain the following expressions:
\ba\label{PB1-func} \{ u_{n,i},u_{m,j}\}_1&=&(-1)^j(\delta^-_{n,K}+\delta^-_{m,K}-1)\nn \\
&&
(u_{n+m-K,i} \delta_{i,j+n-K}-(-1)^{(m+K)(n+K+1)}u_{n+m-K,j} \delta_{i,j-m+K}),\nn\\
\{ u_{n,i},v_{m,j} \}_1&=&(-1)^j\biggl[(\delta^-_{n,K}-1)
(v_{m-n+K,i} \delta_{i,j+n-K}-(-1)^{(m+M)(n+K+1)}v_{m-n+K,j}
\delta_{i,j+m-M}) \nn\\
&&-\delta^+_{m,M}
(u_{n-m+M,i} \delta_{i,j+n-K}-(-1)^{(m+M)(n+K+1)}u_{n-m+M,j} \delta_{i,j+m-M})\biggr],\nn\\
\{ v_{n,i},v_{m,j}\}_1&=&(-1)^j(1-\delta^+_{n,M}-\delta^+_{m,M})
\nn\\ &&(v_{n+m-M,i}
\delta_{i,j-n+M}-(-1)^{(m+M)(n+M+1)}u_{n+m-M,j}\delta_{i,j+m-M})
\ea
for the first Hamiltonian structure and
%
\ba\label{PB2-func} &&\{ u_{n,i},u_{m,j} \}_2=-(-1)^j\frac12
\Bigr[ u_{n,i} u_{m,j}( \delta_{i,j+n-K}
-(-1)^m   \delta_{i,j-m+K})  \nn  \\
&&~~~~+\sum\limits_{k=0}^{n+m} (\delta^+_{m,k}-\delta^-_{m,k})
\Bigr((-1)^{mk}u_{n+m-k,i} u_{k,j} \delta_{i,j+n-k}
-(-1)^{m(n+k+1)}u_{k,i} u_{n+m-k,j}\delta_{i,j-m+k}\Bigl)\Bigr],\nn\\
&&\{ u_{n,i},v_{m,j} \}_2=-(-1)^j\frac12
 \Bigr[ u_{n,i}  v_{m,j}( \delta_{i,j}+
   \delta_{i,j+n-K}
-(-1)^m (  \delta_{i,j+m-M} +   \delta_{i,j+n+m-K-M}))
  \nn \\
&&~~~~+2\sum\limits_{k=\mbox{\footnotesize max}(0,m-n)}^{m-1}
\Bigr(u_{n-m+k,i} v_{k,j+m-k} \delta_{i,j+n-K}
-(-1)^{(n+1)m} v_{k,j} u_{n-m+k,i+k-m}
\delta_{i,j+m+M}\Bigl)\Bigr], \nn\\
&&\{ v_{n,i},v_{m,j} \}_2=(-1)^j\frac12
 \Bigr[ v_{n,i} v_{m,j}( \delta_{i,j-n+M}
-(-1)^m   \delta_{i,j+m-M})
  \nn\\
&&~~~~-\sum\limits_{k=0}^{n+m} (\delta^+_{m,k}-\delta^-_{m,k})
\Bigr((-1)^{mk}v_{n+m-k,i} v_{k,j} \delta_{i,j-n+k}
-(-1)^{m(n+k+1)}v_{k,i} v_{n+m-k,j}
\delta_{i,j+m-k}\Bigl)\Bigr]\nn \ea
for the second Hamiltonian structure where
$$
\delta_{n,m}^+ =\left\{
   \begin{array}{rcl}
1,& \mbox{if} &n>m\\
0,& \mbox{if} &n\leq m \\
 \end{array}
\right. \ , \hspace*{1cm} \mbox{}
\delta_{n,m}^- =\left\{
   \begin{array}{rcl}
1,& \mbox{if} &n<m\\
0,& \mbox{if} &n\geq m. \\
 \end{array}
\right.
$$
Let us remind  that the second Hamiltonian structure  is valid for
even values of $(K,M)$ only.

The Hamiltonian structures thus obtained possess  the properties
(\ref{sym}--\ref{Jac}) with $d_\mathbb{{O}}=d_{L^+_K}=d_{L^-_M}$.
Using them one can rewrite flows (\ref{Eqs1}) for even values of
$(K,M)$ in the bi-Hamiltonian form
\ba D^\pm_s \Biggr( \begin{array}{c} u_{n,i}^{(r)}\\
v_{n,i}^{(r)}\end{array}\Biggr) =\{\Biggr( \begin{array}{c} u_{n,i}^{(r)}\\
v_{n,i}^{(r)}\end{array}\Biggr),H^\pm_{s+1}\}_1=\{\Biggr( \begin{array}{c} u_{n,i}^{(r)}\\
v_{n,i}^{(r)}\end{array}\Biggr),H^\pm_{s}\}_2,\nn\ \ \ \ \ea
with  Hamiltonians
\ba \label{ham2D}
H^+_s=\frac1sstr(L^+_K)^s_*=\frac1s\sum_{i=-\infty}^\infty
(-1)^iu_{sK,i}^{(s)},\ \ \ \ \ \ \
H^-_s=\frac1sstr(L^-_M)^s_*=\frac1s\sum_{i=-\infty}^\infty
(-1)^iv_{sM,i}^{(s)}.\ea
For odd values  of $(K,M)$ one can reproduce  the bosonic flows of
(\ref{Eqs1}) only. In this case eqs. \p{ham2D}, due to relation
\p{evenH}, give only fermionic nonzero Hamiltonians using which
the bosonic flows can be generated via odd first Hamiltonian
structure \p{PB1-func}
\begin{eqnarray} D^\pm_{2s} \Biggr( \begin{array}{c} u_{n,i}^{(r)}\\
v_{n,i}^{(r)}\end{array}\Biggr) =\{\Biggr( \begin{array}{c} u_{n,i}^{(r)}\\
v_{n,i}^{(r)}\end{array}\Biggr),H^\pm_{2s+1}\}_1.\nn\ \ \ \ \ea

\section{Conclusion}

 In this paper, we have generalized the $R$-matrix method
to the case of $Z_2$-graded operators with an involution and found
that there exist two Poisson bracket structures.  The first
Poisson bracket is defined for both odd and even operators with
$Z_2$-grading while the second one is found for even operators
only. It was shown that properties of the Poisson brackets were
provided by the properties of the generalized graded bracket.
 Then we have proposed the Lax-pair representation in terms of the
generalized graded bracket of the new 2D fermionic $(K,M)$-Toda
lattice hierarchy and  applied the developed R-matrix formalism to
derive its bi-Hamiltonian srtucture.  For even values of $(K,M)$
both even first and second Hamiltonian structures were obtained
and for this case all the flows of the 2D fermionic $(K,M)$-TL
hierarchy can be rewritten in
 a bi-Hamiltonian form. For odd values of $(K,M)$
 odd first Hamiltonian structure was found and  for this case
only bosonic flows of the 2D fermionic $(K,M)$-TL hierarchy can be
represented in a Hamiltonian form using fermionic Hamiltonians.

Thus, the problem of Hamiltonian description of the fermionic
flows of the 2D fermionic $(K,M)$-TL hierarchy is still open.
Other problems  yet  to be answered are the construction of the
second Hamiltonian structure (if any) for odd Lax operators and of
the Hamiltonian structures (if any) for Lax operators $L_K^+$ and
$L_M^-$ of opposite  $Z_2$-parities.  Last but not least is the
question of interrelation between  the graded modified Yang-Baxter
equation \p{YB} proposed in this paper and the  graded classical
Yang-Baxter equation introduced in the pioneer paper \cite{KulSk}.
All these questions are a subject for future investigations.

\vspace{.8cm}
 {\bf Acknowledgments.  } We would like to thank A.P. Isaev, P.P.
 Kulish, and  A.A. Vladimirov
 for useful discussions.  This work was
partially supported by  RFBR-DFG Grant No. 04-02-04002, DFG Grant
436 RUS 113/669-2, the NATO Grant PST.GLG.980302, and by the
Heisenberg-Landau program.


\begin{thebibliography}{99}

\bibitem{Mikh} A.V. Mikhailov, Pisma Zh. Eksp. Teor. Fiz. {\bf 30}
(1979) 443.

\bibitem{UT} K. Ueno and K. Takasaki,
Adv. Stud. in Pure Math. {\bf 4} (1984) 1.

\bibitem{Olsh} M.A. Olshanetsky, Commun. Math. Phys. {\bf 88}
(1983) 63.

\bibitem{LSS}  D.A. Leites, M.V. Saveliev, and V.V. Serganova,
"Embeddings of $osp(1|2)$ and the associated nonlinear
supersymmetric equations", in Group Theoretical Methods in
Physics, Vol. {\bf I} (Yurmala, 1985), VNU Sci. Press, Utrecht,
1986, 255.

\bibitem{And} V.A. Andreev, Theor. Math. Phys. {\bf 72} (1987)
758.

 \bibitem{Ik} K. Ikeda,  Lett. Math. Phys. {\bf 14} (1987) 321.

\bibitem{EH} J. Evans and T. Hollowood,
 Nucl. Phys. B {\bf 352} (1991) 723.



 \bibitem{LSor1}
O. Lechtenfeld and A. Sorin, Nucl. Phys.
 B {\bf 557} (1999) 535;  J. Nonlin. Math. Phys. {\bf 8} (2001)
 183; J. Nonlin. Math. Phys. {\bf 11} (2004)
 294.

\bibitem{KS1}
V.G. Kadyshevsky and A.S. Sorin, "Supersymmetric Toda lattice
hierarchies", In "Integrable Hierarchies and Modern Physical
Theories" (Eds. H. Aratyn and A.S. Sorin), Kluwer Acad. Publ.,
Dordrecht/Boston/London, (2001) 289, nlin.SI/0011009.

\bibitem{CzJ} V.V.Gribanov, V.G. Kadyshevsky, and A.S. Sorin,
 Czech. J. Phys. {\bf 54}  (2004) 1289.

 \bibitem{DDNS} V.V.Gribanov, V.G. Kadyshevsky, and A.S. Sorin,  Discrete Dynamics in Nature and Society
{\bf 2004:1} (2004) 113.

\bibitem{DLS}  V.B. Derjagin, A.N. Leznov, and A. Sorin,
   Nucl. Phys. B {\bf 527} (1998)
  643.

\bibitem{Carlet} G. Carlet,
 math-ph/0403049.

\bibitem{KS2}
V.G. Kadyshevsky and A.S. Sorin,  Theor. Math. Phys. {\bf 132}
(2002) 1080; "Continuum limit of the $N=(1|1)$ supersymmetric Toda
lattice hierarchy", JHEP Proceedings, PrHEP unesp2002, Workshop on
Integrable Theories, Solitons and Duality, 1-6 July 2002, Sao
Paulo, Brazil.

\bibitem{AB} H. Aratyn and K. Bering,
 nlin.SI/0402014.


\bibitem{STSh} M.A. Semenov-Tyan-Shanskii,  Functsional. Anal. i Prilozhen. {\bf 17} (1983) 17
(in Russian); Functional Anal. Appl. {\bf 17} (1983) 259.

\bibitem{Yung} C.M. Yung, Mod. Phys. Lett. {\bf 8} (1993) 129.


\bibitem{OR} W. Oevel and O. Ragnisco,  Phys.A {\bf 161} (1989)
181.

\bibitem{LCP} L.Ch. Li  and S. Parmenter, Comm. Math. Phys. {\bf 125} (1989) 545.
\nopagebreak[4]

\bibitem{KulSk}
P.P. Kulish and  E.K. Sklyanin,  J. Sov. Math. {\bf 19} (1980)
1596; Zap. Nauch. Sem. LOMI {\bf 95} (1980) 129 (in Russian).
\end{thebibliography}
\end{document}